\documentclass[useAMS,usenatbib]{mn2e}

\usepackage[dvipdfmx]{graphicx}

\title[Simulating Ly$\alpha$ emitters]
{Lyman-$\alpha$ Emitters in Cosmological Simulations I: Lyman-$\alpha$ Escape Fraction
and Statistical Properties at $z=3.1$} 
\author[Shimizu, Yoshida and Okamoto]{Ikkoh Shimizu,$^{1}$\thanks {E-mail:ikko.shimizu@ipmu.jp}
Naoki Yoshida$^{1}$ \thanks {E-mail:naoki.yoshida@ipmu.jp}, 
Takashi Okamoto$^{2}$ \thanks {E-mail:tokamoto@ccs.tsukuba.ac.jp} \\
$^{1}$Institute for the Physics and Mathematics of the Universe, TODIAS, \\
The University of Tokyo, 5-1-5 Kashiwanoha, Kashiwa, Chiba 277-8583, Japan \\
$^{2}$Center for Computational Sciences, University of Tsukuba, Tsukuba 305-8577, Japan}
\begin{document}

\date{In original form 2011 February 1}

\pagerange{\pageref{firstpage}--\pageref{lastpage}} \pubyear{2010}

\maketitle

\label{firstpage}

\begin{abstract}
We use very large cosmological Smoothed-Particle-Hydrodynamics simulations
to study the properties of high redshift Lyman-$\alpha$ emitters (LAEs).
We identify star-forming galaxies at $z=3.1$ in a
cosmological volume of 100 $h^{-1}$Mpc on a side.
We develop a phenomenological model of absorption, scattering and escape
of Lyman-$\alpha$ photons on the assumption that 
the clumpiness of the inter-stellar medium in a galaxy 
is correlated with the larger scale substructure richness. 
The radiative transfer effect proposed by \citet{Neufeld1991}
allows a large fraction of Lyman-$\alpha$ photons to escape 
from a clumpy galaxy even if it contains a substantial amount of dust.
Our model reproduces, for the first time, all of the following 
observed properties of LAEs at $z=3.1$: the angular correlation function,
ultra-violet and Lyman-$\alpha$ luminosity functions, 
and the equivalent width distribution. 
A simple model that takes only dust absorption into account
fails in matching the observational data, suggesting that the kind
of effect we consider is needed. Our model also predicts
a bimodal age distribution for LAEs. There are old, massive
and dusty LAEs, similar to recently found high redshift LAEs.
The large LAEs have escape fractions of Lyman-$\alpha$ photons
of $f_{\rm esc} \sim 0.05-0.1$.
\end{abstract}


\section{Introduction}
A population of star-forming galaxies at high redshifts are characterized
by their strong Lyman-$\alpha$ (Ly$\alpha$) line emission.
Such Ly$\alpha$ emitters (LAEs) have been found at various redshifts
by narrow-band surveys using 8-10~m class telescopes 
\citep{Hu98, Hu99, Hu02,Kodaira03, Shimasaku03, Shimasaku06, 
Ha2004, Ou04, Ou05, Ouchi08, Taniguchi05, Matsuda04, Matsuda05, Iye06}. 

It is generally thought that the strong Ly$\alpha$ emission physically  
originates from star-forming regions (H{\sc ii} regions) 
in a young starburst galaxy.
Some LAEs have very large equivalent widths (${\rm EW_{Ly\alpha}}$) exceeding $400 {\rm \AA}$, 
which is difficult to explain with ordinary stellar population synthesis models 
(e.g., \citet{CF93, Schaerer03}). 
Alternative physical models include cooling radiation from a primordial collapsing 
gas \citep{Haiman00, Fardal01}, from a galactic wind-driven shell \citep{TS00},
and from supernova remnants \citep{MUF04, MU2006}.  

Recent large LAE surveys provided an array of statistical properties of LAEs 
such as the Ly$\alpha$ luminosity function, two-point angular correlation 
function and the evolution of them. The observations generally suggest that 
LAEs are not simply a subset of star-forming galaxies.
Indeed, theoretical models proposed so far do not fully explain 
the observed properties.
In particular, reproducing very large equivalent widths of some
bright LAEs appears to be challenging. 
Ly$\alpha$ photons are easily absorbed by dust and thus
it is naively expected that LAE is a very young and dust-free galaxy. 
While some observations and theoretical studies actually support the notion
\citep{Ga2006, Ga2007, MU2006, S2007},
more recent multi-wavelength observations of LAEs 
in optical, infrared and sub-millimeter suggest that 
there are LAEs that are indeed old and dusty 
\citep{Fin2007, Lai2008, Matsuda2007, Uchimoto2008, Fin2009c, SMGLAE, Ono2010}. 
Interestingly, such a population increases with decreasing redshift \citep{Nilsson2009}. 
There is even an evidence that some sub-millimeter galaxies show 
strong Ly$\alpha$ emission \citep{Smail04}. 
The existence of a substantial amount of dust appears incompatible 
with strong Ly$\alpha$ emission. There must be a physical mechanism
for Ly$\alpha$ photons to escape from such dusty galaxies.
 
There have been a number of theoretical studies on the population of
LAEs. \citet{Nagamine10} study a stochastic model
where a galaxy goes through LAE phase occasionally.
\citet{Shimizu2010} argue that an old evolved galaxy can become a LAE 
as a consequence of delayed gas accretion.
Their model assumes that the delayed starburst occurs at outskirt 
of a galaxy so that Ly$\alpha$ photons can escape easier 
than those emitted from the central region. 
\citet{Dayal2009, Dayal2010, Dayal2011} perform cosmological simulations, 
which reproduce well the UV and Ly$\alpha$ luminosity functions. 
\citet{Dayal2011} also calculate the neutral hydrogen fraction at $z = 5.7$ 
by using a combination of their LAE formation model and radiation transfer 
calculation of hydrogen reionization. 
In the most recent work of \citet{Dayal2011}, 
absorption of Ly$\alpha$ photons by dust 
is treated in a very simple manner, where a dusty gas is a pure absorber.

It has been suggested that large-scale gas motions in and around galaxies 
can affect the absorption of Ly$\alpha$ photons \citep{Zheng2010a, Zheng2010b}. 
Strong galactic winds from Ly$\alpha$ emitting 
galaxies are indeed found in the local universe 
\citep{Lequeux1995, Kunth1998, Kunth2003, Mas-Hesse2003, Keel2005}. 
Strong winds are also seen in high-$z$ Lyman break 
galaxies (LBGs) with strong Ly$\alpha$ emission 
\citep{Pettini2002, Shapley2006, Bower2004, Wilman2005, Frye2007, 
Pentericci2007, Tapken2007}. 
Although the large-scale velocity structure is rather important,
it is unlikely that all the LAEs blow strong galactic winds
because only a small fraction of high redshift galaxies shows the signature
of strong outflow \citep{McLinden2010}. 
It appears that another physical process is necessary for massive
dusty galaxies to be bright LAEs.

\citet{Neufeld1991} proposed an important effect for Ly$\alpha$ transfer.
In a clumpy, multi-phase inter-stellar medium (ISM),
dust is locked up in small cold clouds. 
Ly$\alpha$ photons can then escape from the clumpy ISM {\it easier} than 
continuum photons because Ly$\alpha$ photons, having a very large 
scattering cross-section, are preferentially scattered 
at the surface of the clouds.
On the other hand, UV continuum photons are easily absorbed by dust. 
Neufeld's model can explain not only the existence of Ly$\alpha$ emission 
from dusty galaxies but also the observed high equivalent widths 
\citep{HansenOh2006, Kobayashi2007, Kobayashi2010, Fin2008, Fin2009a, Fin2009b}. 
Clearly it is important to incorporate the effect in modelling LAEs.

In this paper, we study the statistical properties of LAEs at $z=3.1$.
We perform large cosmological hydrodynamic simulations for the
standard $\Lambda$CDM cosmology. 
Our simulations follow star formation, supernova feedback, 
and metal enrichment. For galaxies identified in our cosmological 
simulation,
we calculate the spectral evolution and dust extinction. 
Unlike in semi-analytic methods, we can directly study
the internal structure of simulated galaxies 
as well as their spatial distribution in a cosmological volume.

Throughout this paper, 
we adopt the $\Lambda$CDM cosmology with the matter density $\Omega_{\rm{M}} = 0.27$, 
the cosmological constant $\Omega_{\Lambda} = 0.73$, 
the Hubble constant $h = 0.7$ in units of $H_0 = 100 {\rm ~km ~s^{-1} ~Mpc^{-1}}$, 
the baryon density $\Omega_{\rm B} = 0.046$, 
and the matter density fluctuations are normalized by setting
$\sigma_8 = 0.81$ \citep{WMAP}. 
All magnitudes are expressed in the AB system, 
and all Ly$\alpha$ EW$_{\rm Ly\alpha}$ values in this paper are 
in the rest frame. 

\section{Theoretical Model}
\subsection{Numerical simulations}
Our simulation code is based on an early version of the Tree-PM 
smoothed particle hydrodynamics (SPH) code {\scriptsize GADGET-3} which 
is a successor of Tree-PM SPH code {\scriptsize GADGET-2} \citep{Gadget}.   
We simulate $N = 2 \times 640^3$ particles in a comoving volume of 
100 $h^{-1}{\rm ~Mpc}$ on a side. 
The mass of a dark matter particle and that of a gas particle 
are $2.41 \times 10^8 h^{-1}{\rm M_{\odot}}$ and $4.95 \times 10^7 h^{-1}{\rm M_{\odot}}$, 
respectively.

Physical processes such as star formation and feedback are implemented
as in \citet{Okamoto2008, Okamoto2009, Okamoto2010}. 
In particular, our simulations employ a new galactic wind model 
in which the initial velocity of a wind particle is proportional to 
the local velocity dispersion of the dark matter particles.
This is motivated by observations that suggest large scale 
outflows have velocities that scale with the circular velocity 
of their host galaxies \citep{Martin2005}.  
As a proxy for host halo's circular velocity, which is not easily 
calculated on-the-fly, we use the local one-dimensional velocity 
dispersion, determined from neighbouring dark matter particles. 
\citet{Okamoto2010} found that this quantity, $\sigma$, is strongly 
correlated with the maximum circular velocity of host (sub-) halos,
$v_{\rm max}$, and the relation between these quantities does not 
evolve with redshift. 
This prescription results in a wind speed that increases as a halo 
grows and hence a wind mass-loading (wind mass per unit star formation 
rate) is highest at early times (or in small halos). 
This scaling has been shown to reproduce the physical properties 
of the local group satellites \citep{Okamoto2009, Okamoto2010}. 

Our simulations include the time-evolving photoionization background 
\citep{Haardt2001}, 
metallicity-dependent gas cooling and photoheating \citep{Wiersma2009},
supernovae feedback and chemical enrichment \citep{Okamoto2005, Okamoto2008}.
We use metallicity-dependent stellar lifetimes and chemical yields 
\citep{Portinari1998, Marigo2001}. 
The details of these processes are found in the above references.
Here we give a brief description. 
Each SPH particle can spawn a new star particle when the particle 
satisfies a set of standard criteria for star formation. 
A star particle carries its properties such as mass, formation time and metallicity. 
We calculate the spectral energy distribution (SED) of each star particle 
using the population synthesis code {\scriptsize P\'{E}GASE} \citep{PEGASE}.
We then sum the individual SEDs to obtain the total SED of a simulated galaxy. 

We run a friends-of-friends group finder \citep{FOF}
to locate groups of stars, i.e., galaxies.
We also identify substructures (subhalos) in each FoF group 
using SUBFIND algorithm developed by \citet{Springel2001}. 
For the identified galaxies, we calculate the intrinsic Ly$\alpha$ luminosities 
using 'PEGASE'. 
In PEGASE, two thirds of ionizing photons are converted to Ly$\alpha$ 
photons under the assumption of the case B recombination. 
Finally, we calculate the effect of dust extinction 
on ultra-violet continuum and Ly$\alpha$ emission. 
This is the key component of our model.
We will describe the details in the next subsection. 

\subsection{Model description for dust absorption}
We calculate the SED and the Ly$\alpha$ luminosity
for individual galaxies identified in our cosmological simulation. 
In order to compare our model predictions directly with observational data, 
we need to include the effect of dust absorption. 
We first assume that the dust mass to metal mass ratio is 
a constant of $0.4$, consistent with the local value.
Note that the metallicity of a galaxy is calculated from 
the metallicities of gas particles. 
We further assume that the galaxies are roughly spherical;
we use only one length scale, the effective radius, to evaluate the optical depth.
We calculate the optical depth $\tau_{\rm d}(\lambda)$ 
for UV continuum photons as
\begin{equation}
\tau_d(\lambda) = \frac{3 \Sigma_d}{4a_{\rm d} s},
\end{equation}
where $a_{\rm d}$ and $s$ are the typical size of dust grains, 
and the material density of dust grains, respectively. 
We adopt the standard choice of $a_{\rm d} = 0.1 ~{\rm \mu m}$ and $s = 2.5 ~{\rm g}
~{\rm cm}^{-3}$ \citep{Todini2001, Nozawa2003}.
The dust surface mass density $\Sigma_d$ is 
\begin{equation}
\Sigma_d = \frac{M_{\rm d}}{\pi r_{\rm d}^2}, 
\end{equation}
where $M_{\rm d}$ and $r_{\rm d}$ are the total dust mass ($40\%$ of metal mass) 
and the effective radius of the galaxy, respectively.  
The effective radius $r_{\rm d}$ is a fraction of the virial radius
and is given by $f_{\rm d} r_{\rm vir}$ with $f_{\rm d} = 0.18$.
The escape fraction of UV continuum photons $f_{\rm cont}(\lambda)$ is 
then calculated as 
\begin{equation}
f_{\rm cont}(\lambda) = \frac{1 - \exp{(- \tau_{\rm d}(\lambda))}}{\tau_{\rm d}(\lambda)}.
\end{equation}
We use the dust optical constant $Q$ as a function of wavelength 
given in \citet{Draine1984}.
We also calculate the IGM absorption at the blue side of $1216 \rm \AA$ 
following \citet{Madau1995}.

A crucial quantity in our model 
is the effective optical depth of Ly$\alpha$ 
line, $\tau_{\rm Ly\alpha}$.  
We employ two models to calculate the effective optical depth.
A simplest assumption would be 
to set 
\begin{equation}
\tau_{\rm Ly\alpha}^{\rm abs} = c_{\rm abs} \tau_{\rm d},
\label{eq:tau_abs}
\end{equation}
where $c_{\rm abs}$ is a constant parameter. 
Then, the escape fraction of Ly$\alpha$ photons $f_{{\rm Ly\alpha}}^{\rm abs}$ is 
given by
\begin{equation}
f_{{\rm Ly\alpha}}^{\rm abs} = \frac{1 - \exp{(- \tau_{\rm Ly\alpha}^{\rm abs})}}{\tau_{\rm Ly\alpha}^{\rm abs}}. 
\end{equation}
We call this model as pure absorption model. 
Essentially, dust is treated as a pure absorber of Ly$\alpha$ photons
in this model (e.g., \citet{Dayal2009, Dayal2010, Dayal2011}).
Ly$\alpha$ photons can not easily escape from dusty galaxies
which have a large effective optical depth.

The second model, which we propose in the present paper, 
is motivated by the multiphase inter-stellar medium (ISM) model 
of \citet{Neufeld1991}. 
In a clumpy ISM, Ly$\alpha$ photons are scattered mostly at the surface
of cold clumps before they are absorbed by dust. 
A large fraction of Ly$\alpha$ photons can then
escape from a clumpy ISM through multiple scatters.
The key quantity here is the overall clumpiness of the ISM in a galaxy.
Since our cosmological simulation does not resolve the fine structure of the 
ISM, we need to estimate the clumpiness of the ISM in some way.
It is probably reasonable to expect that the ISM structure is well
developed in a massive galaxy which itself has rich substructures.
We make an assumption that the clumpiness of the ISM
has a correlation with the larger scale internal structure of a galaxy. 
In practice, we use the number of subhalos (satellites) $N_{\rm sub}$ 
of the galaxy as a measure of its ``clumpiness''.
We introduce the clumpiness factor $S$, which depends on $N_{\rm sub}$,  
and express the effective optical depth of Ly$\alpha$ line as 
\begin{equation}
\tau_{\rm Ly\alpha}^{\rm sub} = c_{\rm sub} S \tau_{\rm d}, 
\end{equation}
where $c_{\rm sub}$ is a normalization constant. 
For simplicity, we assume the clumpiness factor is expressed as a power-law
\begin{equation}
S = N_{\rm sub}^{\alpha}.
\label{eq:Sclump}
\end{equation} 
After some experiments, we find that setting $\alpha=-0.5$
works remarkably well, reproducing nearly all the important
observational data, as will be shown in the following sections.
We call this model the substructure model. 
The escape fraction of Ly$\alpha$ photons $f_{{\rm Ly\alpha}}^{\rm sub}$ in this model is
given by
\begin{equation}
f_{{\rm Ly\alpha}}^{\rm sub} = \frac{1 - \exp{(-\tau_{\rm Ly\alpha}^{\rm sub})}}{\tau_{\rm Ly\alpha}^{\rm sub}}. 
\label{eq:fesc_sub}
\end{equation}

The combination of equation (\ref{eq:Sclump}) and equation (\ref{eq:fesc_sub})
effectively yields a larger escape fraction for a dusty but ``clumpy'' galaxy. 
Finally, we fix the normalization constants in the two models, 
$c_{\rm abs}$ and $c_{\rm sub}$, by matching the number density of 
simulated LAEs with the observed number density $n_{\rm LAEs} \sim 5 \times 10^{-4} ~[\rm Mpc^{-3}]$ \citep{Ouchi08}.
We identify simulated galaxies 
which satisfy ${\rm EW_{Ly\alpha} > 20{\rm \AA}}$ 
and $L_{\rm Ly\alpha}^{\rm obs} > 1.0 \times 10^{42} {\rm [erg / s]}$ as LAEs.  

\section{Result}
\subsection{The Spatial Distribution of Simulated LAEs}

\begin{figure*}
\includegraphics[width = 160mm]{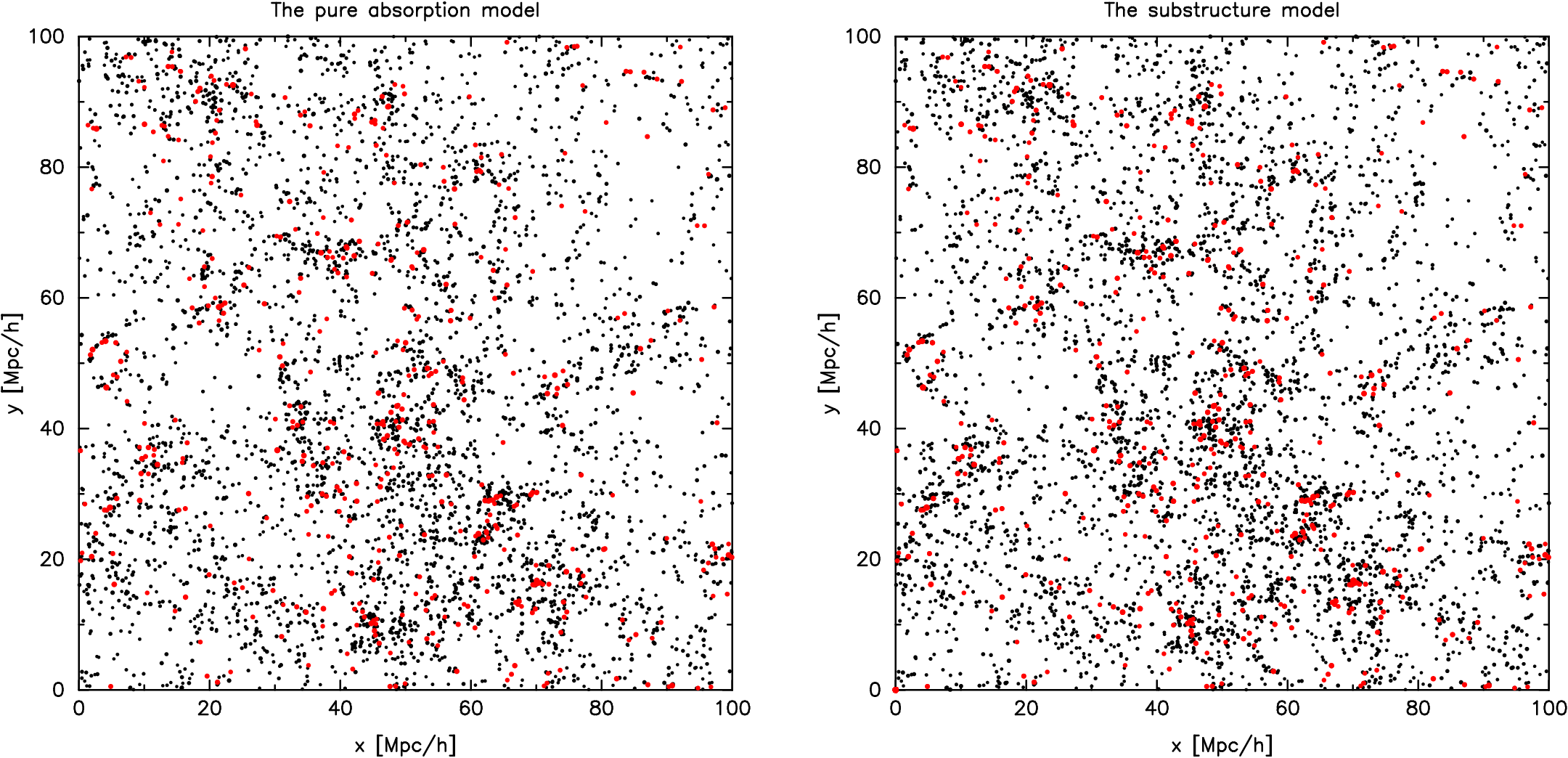}
\caption{The spatial distribution of simulated LAEs. 
The left and right panels show LAEs in the pure absorption model 
and those in the substructure model, respectively. 
The point size is scaled with Ly$\alpha$ luminosity of each galaxy
so that luminous galaxies appear as large points. 
The smallest and biggest points correspond to LAEs with $10^{42} \leq L_{\rm Ly\alpha} \leq 10^{42.2}
{\rm erg}{\rm s}^{-1}$ 
and $L_{\rm Ly\alpha} \geq 10^{43} {\rm erg}{\rm s}^{-1}$, respectively. 
The red-points shows brightest LAEs ($> 6\times10^{42} {\rm ~[erg/s]}$). }
\label{LAEs_dist}
\end{figure*}

\begin{figure}
\includegraphics[width = 80mm]{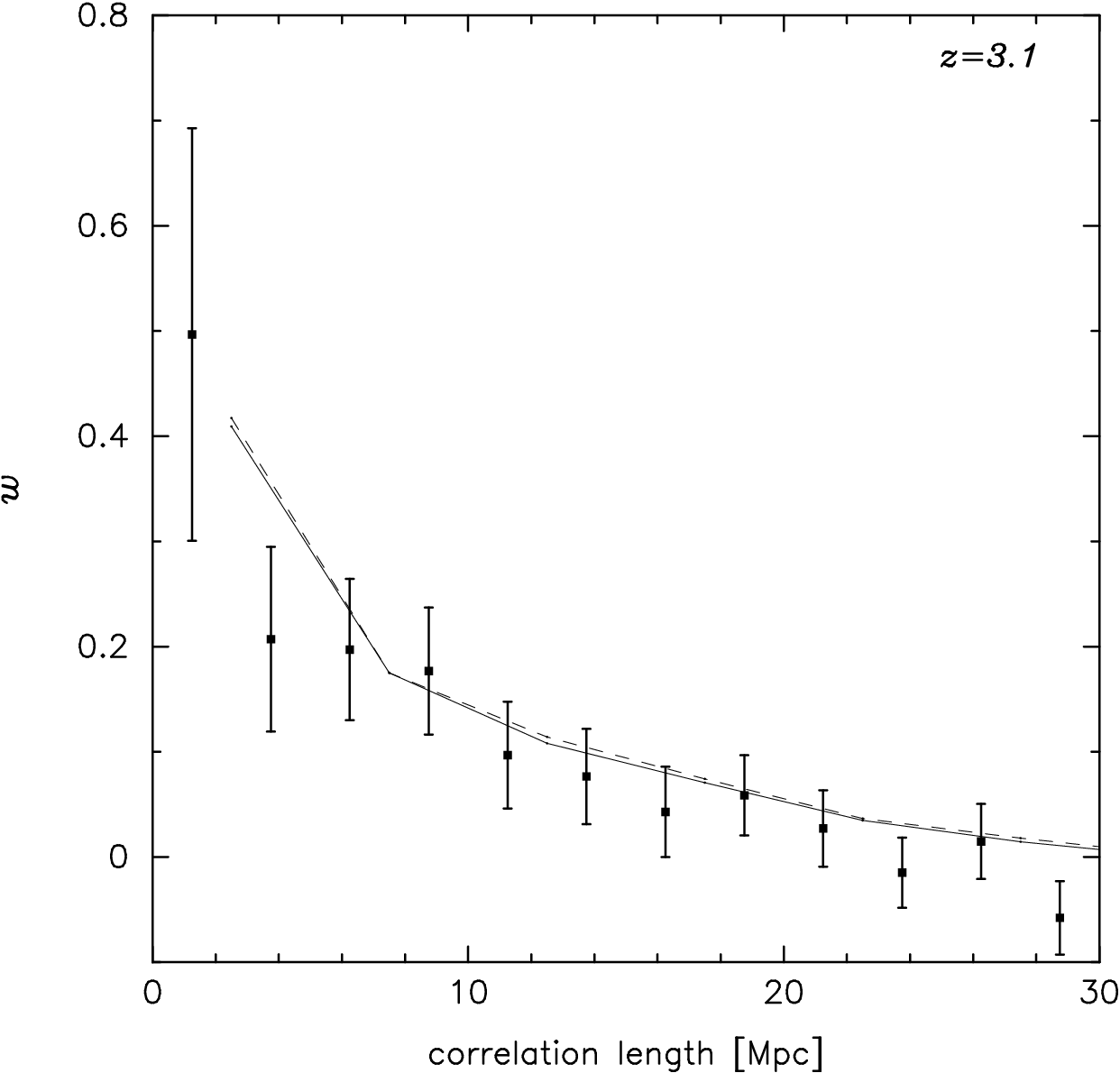}
\caption{The two-point angular correlation function (ACF). 
The solid and dashed lines represent the substructure model 
and the pure absorption model, respectively. 
Points with error bar are the ACF of LAEs observed in SSA22a field \citep{Ha2004}.}
\label{ACF}
\end{figure}

Fig. \ref{LAEs_dist} shows the projected distribution of 
the LAEs at $z=3.1$ in the simulation volume of 100 comoving $h^{-1}$Mpc on a side. 
The point size represents the Ly$\alpha$ luminosity of each galaxy. 
Luminous LAEs are shown by large and red points.
The overall distribution appears quite similar in the two models,
with LAEs approximately tracing the underlying matter distribution.
The luminous LAEs are clustered more strongly in the substructure model, 
although the difference is small.

We quantify the spatial distribution of the simulated LAEs.
We calculate the two-point angular correlation function (ACF) 
and compare it with the observed correlation function.
The ACF of LAEs found in the SSA22 field \citep{Ha2004} is used
for the comparison. 
Fig. \ref{ACF} shows the ACFs for the two models.
Both of the models reproduce the observational result well.
The LAEs are hosted by dark halos with mass of $\sim 10^{11} {\rm M_{\odot}}$. 

\subsection{Ly$\alpha$ Luminosity Functions}
\begin{figure*}
\includegraphics[width = 160mm]{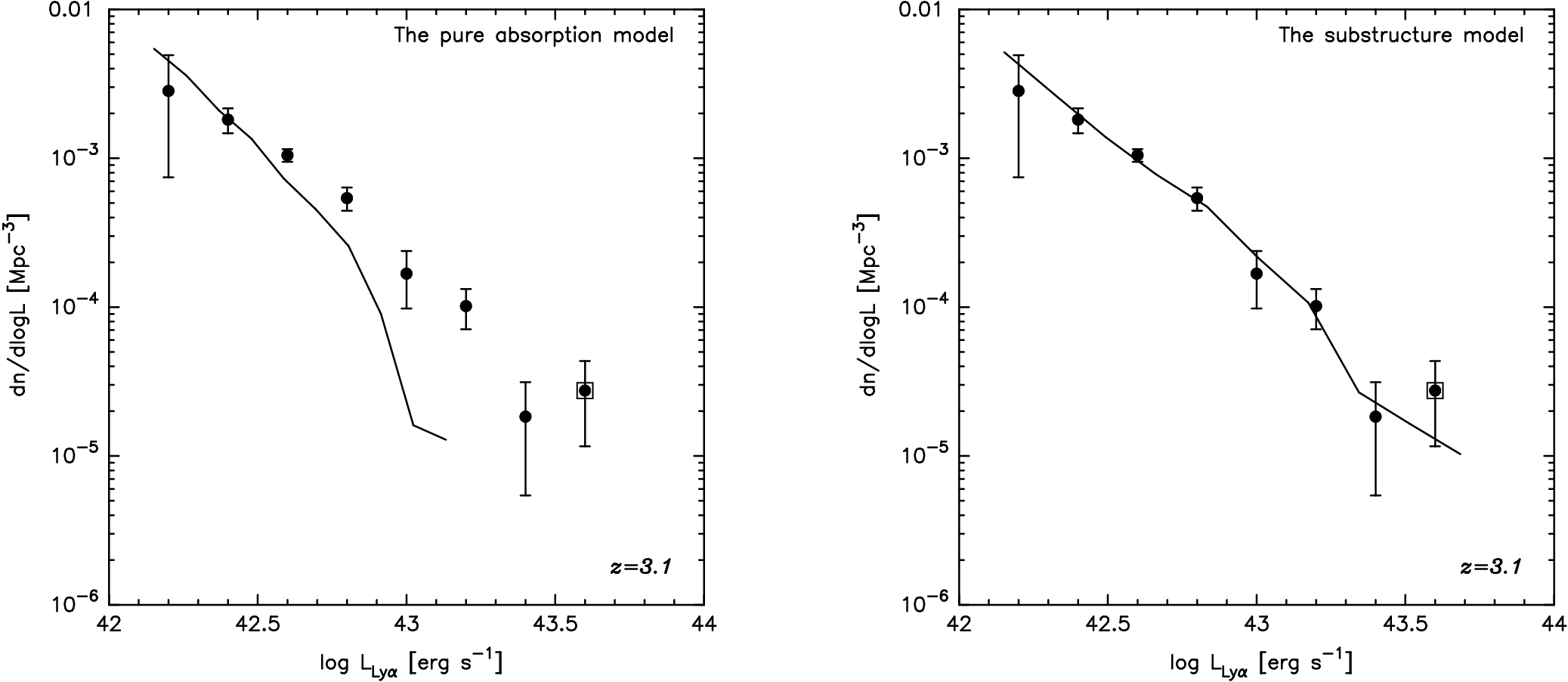}
\caption{The Ly$\alpha$ luminosity function for 
the pure absorption model (left)
and for the substructure model (right).
We compare the model predictions with 
the observational data of \citet{Ouchi08} with error bars. 
The filled circle marked an open square indicates LAE with AGN.}
\label{Lya_LF}
\end{figure*}

In Fig. \ref{Lya_LF}, we compare the Ly$\alpha$ luminosity functions 
with the observational data at $z = 3.1$ \citep{Ouchi08}. 
The pure absorption model (left panel) does not reproduce the observational data
at the bright end. It predicts a smaller number of LAEs
by a factor of ten at $L_{\rm Ly\alpha} \sim 10^{43} {\rm erg}{s}^{-1}$. 
In both our models,
the intrinsic Ly$\alpha$ luminosity of a galaxy 
is proportional to its star formation rate (SFR),
and thus the {\it intrinsically} bright LAEs are hosted typically 
by massive halos. However, such massive galaxies are also aged and dusty. 
Because of the strong dust absorption of Ly$\alpha$ photons 
(see equation [\ref{eq:tau_abs}]), 
dusty star-forming galaxies do not appear as LAEs in the pure
absorption model.

Our substructure model, on the other hand, shows a substantially 
better agreement with the observational data. 
In the substructure model, dusty, aged and massive galaxies have 
complex internal structures.
We assume that the evolved galaxies have also a more complex ISM structure,
where Ly$\alpha$ photons can escape easily because of the Neufeld effect.
Thus, even aged and dusty galaxies appear as bright LAEs
and the resulting luminosity function matches the observation very well
at the bright-end.

Interestingly, the substructure model predicts that 
the Ly$\alpha$ escape fraction of massive galaxies 
is $f_{\rm esc} \sim 0.05-0.1$, with a substantial dispersion. 
In the pure absorption model, the Ly$\alpha$ escape fraction decreases
proportionally to the halo mass because of the size effect
and also because massive galaxies are more metal-enriched. 


\subsection{${\rm M_{UV} - EW_{Ly{\alpha}}}$ distribution}

\begin{figure*}
\includegraphics[width = 160mm]{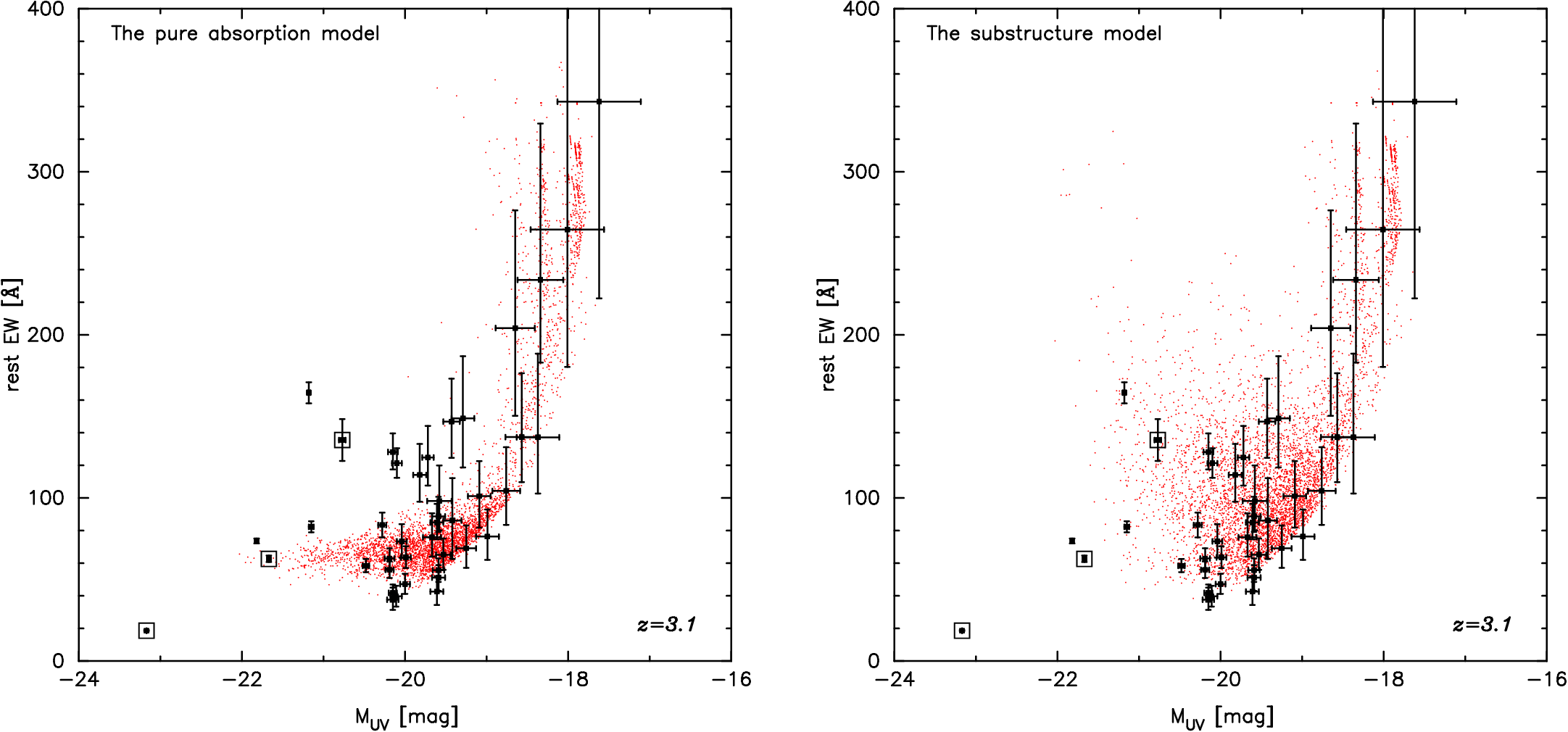}
\caption{Distributions of LAEs in the ${\rm M_{UV} - EW_{Ly\alpha}}$ plane at $z = 3.1$. 
The simulated LAEs are shown by dots, whereas 
points with error bars are observational data of \citet{Ouchi08}. 
The filled circles marked an open square indicate LAEs with AGN. }
\label{Muv_EW}
\end{figure*}

Fig. \ref{Muv_EW} shows the distributions of our simulated 
LAEs in the ${\rm M_{UV} - EW_{Ly\alpha}}$ plane at $z=3.1$. 
In this plot, both models appear similar, with the LAEs populating 
the region where the observational data points also lie. 
Although both the models reproduce the observed trend that
the equivalent width decreases with UV luminosity,
only the substructure model predicts 
the existence of 
UV bright galaxies ($M_{\rm UV} < -20$) with large ${\rm EW_{Ly\alpha}} > 100{\rm \AA}$.
This is again owing to the effect of enhanced Ly$\alpha$ 
luminosity in the substructure model.
We also note here that the substantial dispersion of ${\rm EW_{Ly\alpha}}$
for UV bright LAEs, as is also found in the observations,
is important. Apparently the Lyman$\alpha$ escape fraction is not
a simple function of the galaxy mass only. 
We discuss this issue more in detail in the
discussion section. 

\subsection{Ly$\alpha$ Equivalent Width Distribution}

\begin{figure*}
\includegraphics[width = 160mm]{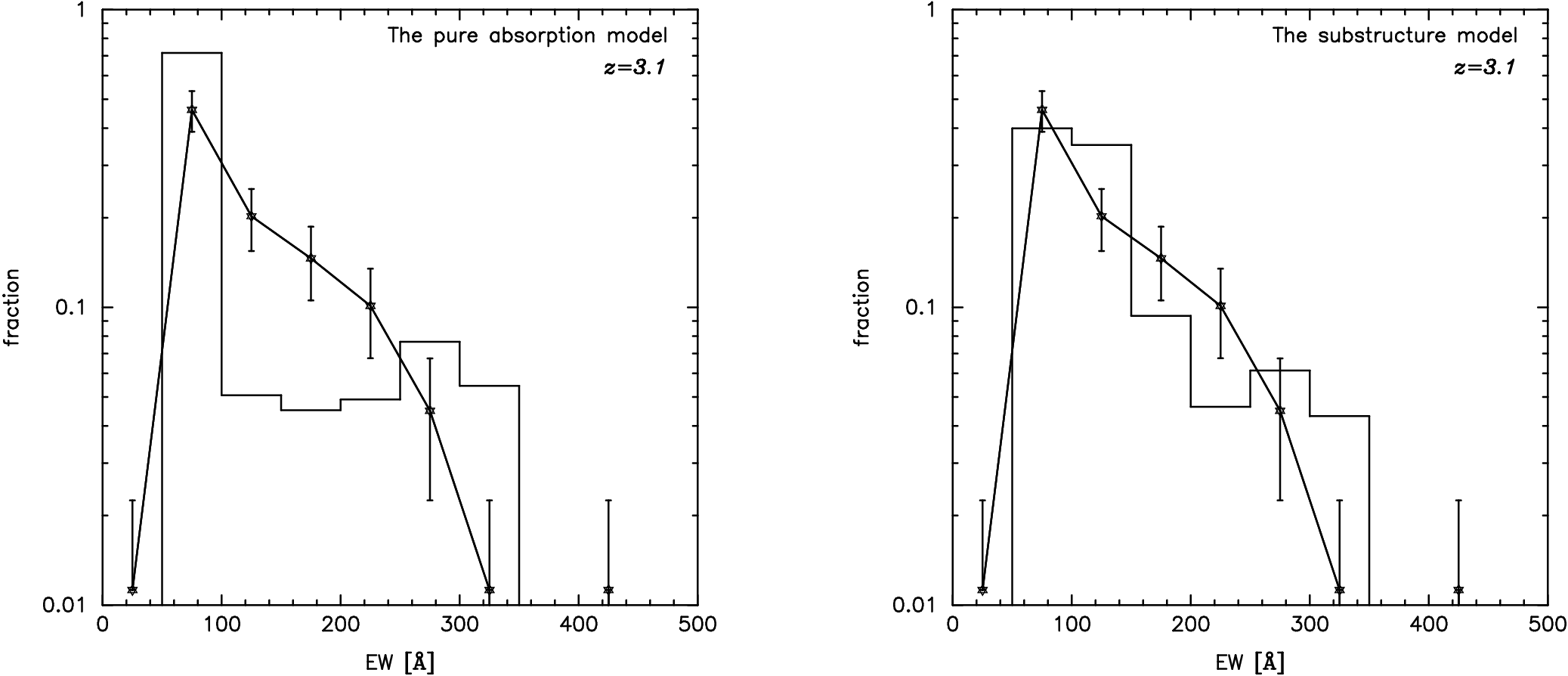}
\caption{The rest-frame Ly$\alpha$ equivalent 
width (${\rm EW_{Ly\alpha}}$) distribution at $z = 3.1$. 
The histogram shows our model results,
whereas the solid lines with error bars show observational results \citep{Ouchi08}.}
\label{EW_dist}
\end{figure*}

In Fig. \ref{EW_dist}, we compare the rest-frame Ly$\alpha$ equivalent 
width of simulated LAEs (histograms) with
the observational data (solid line with error bars). 
In the pure absorption model, ${\rm EW_{Ly\alpha}}$ distribution is 
approximately-constant at ${\rm EW_{Ly\alpha} > 100\AA}$
whereas the observed distribution decreases
toward high ${\rm EW_{Ly\alpha}}$.
The substructure model reproduces 
the observed ${\rm EW_{Ly\alpha}}$ distribution very well. 
It is interesting that the fraction of large ${\rm EW_{Ly\alpha}}$ galaxies 
is similar between the two models.
However, the substructure model predicts more LAEs with 
 ${\rm EW_{Ly\alpha}} \sim 100-200\AA$. For these LAEs, UV continuum
is absorbed by dust significantly but Ly$\alpha$ photons
can escape again by the scattering effect.

\subsection{Age distribution of simulated LAEs}

\begin{figure}
\includegraphics[width = 80mm]{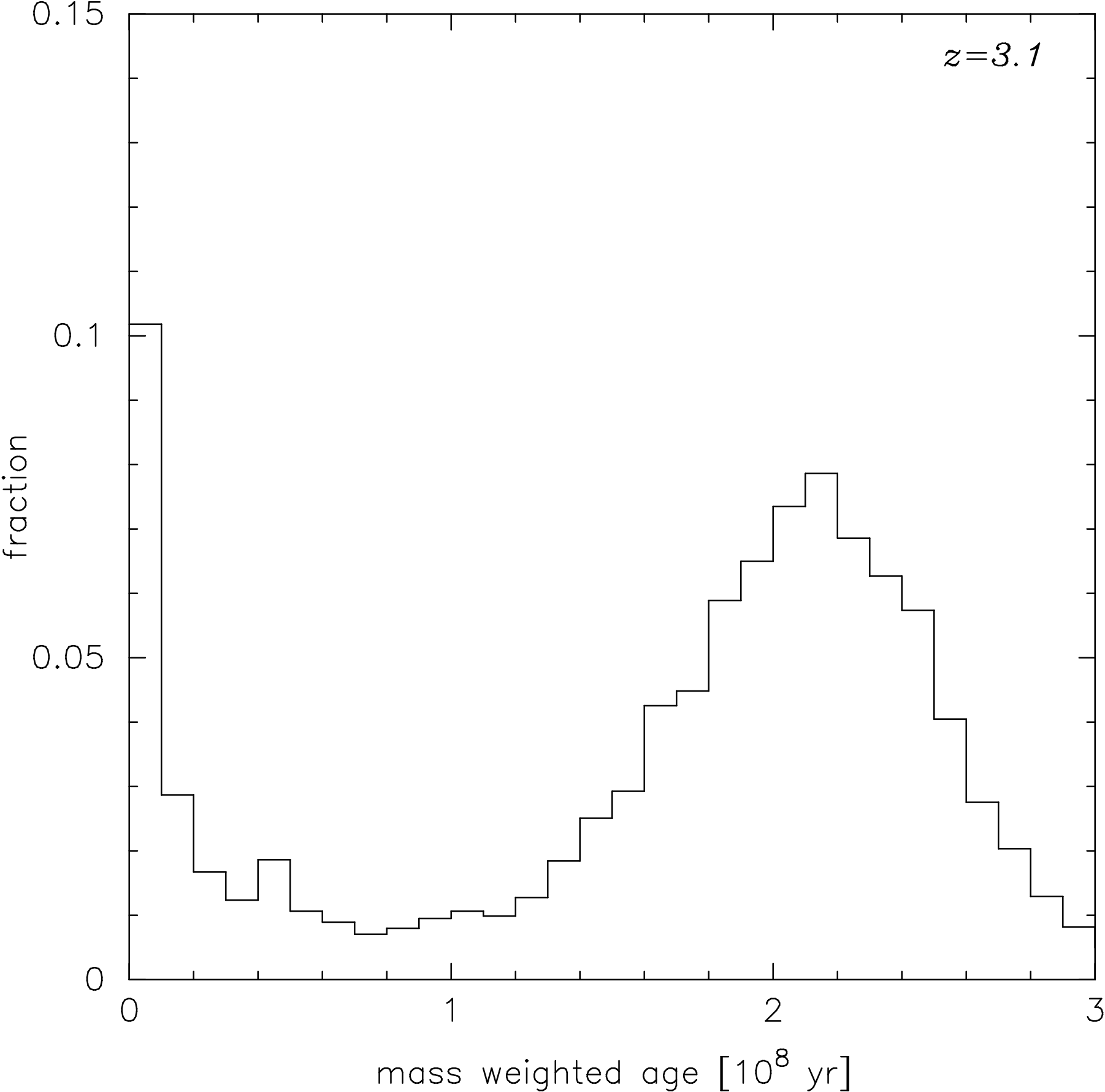}
\caption{The age distribution of simulated LAEs for the substructure model.
There are two types LAEs; young LAEs that are populated at the left of the
plot and old LAEs that are broadly distributed around $t_{\rm age} \sim 2\times 10^{8}$ years.}
\label{AGE}
\end{figure}

Finally, we show the age distribution of simulated LAEs in Fig. \ref{AGE}. 
We assign the age of a simulated LAE by calculating the mass weighted mean
of the stellar ages. 
The simulated LAEs shows clearly a bimodal distribution.
Recently formed galaxies are young LAEs, 
whereas massive, dusty galaxies with
$t_{\rm age} > 10^{8}$ years can also appear as bright LAEs. 
The latter population is indeed found recently 
by \citet{Fin2009b}.

\section{Conclusions and Discussion}
We study a number of observed properties of LAEs using 
a large cosmological hydrodynamic simulation.
The simulation follows the formation and evolution 
of star-forming galaxies by employing new feedback models of \citet{Okamoto2008}.  
We develop a novel model in which Ly$\alpha$ photons can escape from 
a massive, dusty galaxy if the ISM is clumpy.
The idea is motivated by the multiple scattering and escape
of Ly$\alpha$ photons originally proposed by \citet{Neufeld1991}.

Our physical model of LAEs reproduces not only the Ly$\alpha$ luminosity function 
but also ${\rm M_{UV} - EW_{Ly{\alpha}}}$ distribution, the equivalent width distribution 
and the angular two-point correlation function at $z=3.1$. 
It is the first theoretical model that successfully reproduces all these 
observational results using cosmological hydrodynamic simulations. 
Interestingly, the model predicts the Ly$\alpha$ escape fraction is roughly 
constant for the massive LAEs. 
Massive, dusty and aged galaxies can appear as bright LAEs,
as is consistent with recent observations.
Contrastingly, the pure absorption model fails in reproducing
many of the observational results. 
We argue that the kind of effect to enhance Ly$\alpha$ photon escape
from massive dusty galaxies is necessary for modeling LAEs. 

It is interesting to ask if the Ly$\alpha$ escape fraction 
can be described as a simpler function of galaxy mass.
In our simulation, the substructure abundance of simulated 
galaxies roughly scales with their host halo mass, although
with substantial dispersions. 
It has turned out that the dispersion is also important
to reproduce some observational results. 
To explicitly test the idea, we have employed a simpler model 
in which the Ly$\alpha$ escape fraction 
is a function of host halo mass so that the escape fraction
approximately matches to the mean escape
fraction of our substructure model for a given halo mass. 
We have found that this simple model fails in reproducing 
the ${\rm M_{UV} - EW_{Ly{\alpha}}}$ distribution and ACF. 
The simple model boosts the Ly$\alpha$ luminosity too much
and cannot explain the existence of UV-bright LAEs with low equivalent
widths (see Fig. \ref{Muv_EW}). 
Furthermore, in the simple model, massive galaxies 
are identified selectively as LAEs. 
Consequently the strength of ACF becomes large, although 
it is still within the upper bound of the current data.
Overall, we conclude that the internal structure of a galaxy is 
an important factor to determine its appearance as a LAE.
Interestingly, \citet{CF93} argue that a large scatter of 
Ly$\alpha$ emission versus metallicity correlation 
might be caused by the structure of the interstellar medium.

In the present paper, we have focused on the physical properties 
of LAEs at a particular
redshift of $z = 3.1$ where an array of observations of LAEs are available.
According to very recent observations, the fraction of old population 
in observed LAEs increases with decreasing redshift, 
whereas the escape fraction of Ly$\alpha$ emission becomes larger 
at higher redshift \citep{Nilsson2009, Hayes10}. 
It is certainly interesting and important to test
whether or not our model can reproduce not only 
the physical properties of observed LAEs 
but also their evolution.
We will present the properties of simulated LAEs at various 
redshifts in a forthcoming paper (Shimizu et al. 2011 in prep). 
We will also study the epoch of hydrogen reionization using
our theoretical model.

\section*{Acknowledgments}
We are grateful to M. Ouchi, Y. Matsuda and T. Hayashino for providing 
their observational datas.
Numerical simulations have been performed with the EUP and PRIMO 
cluster system installed 
at Institute for the Physics and Mathematics of the Universe, University of Tokyo. 
This work was partially supported by Grant-in-Aid for Young Scientists (S) (20674003).
TO acknowledges the financial support of Grant-in-Aid for Scientific Research (S) (20224002) 
and of Grant-in-Aid for Young Scientists (21840015) by JSPS.

\bsp

\label{lastpage}


\begin{thebibliography}{99}
\bibitem[\protect\citeauthoryear{Bower et al.}{2004}]{Bower2004} Bower, R. G., et al. 2004, MNRAS, 351, 63
\bibitem[\protect\citeauthoryear{Charlot, \& Fall.}{1993}]{CF93} Charlot, S., Fall, S. M., 1993, ApJ, 415, 580
\bibitem[\protect\citeauthoryear{Dayal et al.}{2009}]{Dayal2009} Dayal, P., Ferrara, A., Saro, A., Salvaterra, R., Borgani, S., Tornatore, L. 2009, MNRAS, 400, 2000
\bibitem[\protect\citeauthoryear{Dayal et al.}{2010}]{Dayal2010} Dayal, P., Ferrara, A., Saro, A. 2010, MNRAS, 402, 1449
\bibitem[\protect\citeauthoryear{Dayal et al.}{2011}]{Dayal2011} Dayal, P., Maselli, A., Ferrara, A. 2011, MNRAS, 410, 830
\bibitem[\protect\citeauthoryear{Davis et al.}{1985}]{FOF} Davis, M., Efstathiou, G., Frenk, C. S., White, S. D. M. 1985, ApJ. 292,  371 
\bibitem[\protect\citeauthoryear{Draine \& Lee.}{1984}]{Draine1984} Draine, B. T., Lee, H. M., 1984, ApJ, 285, 89
\bibitem[\protect\citeauthoryear{Fardal et al.}{2001}]{Fardal01} Fardal M.~A., Katz N., Gardner J.~P., Hernquist L., Weinberg D.~H., Dav{\'e} R., 2001, ApJ, 562, 605 
\bibitem[\protect\citeauthoryear{Finkelstein et al.}{2007}]{Fin2007} Finkelstein, S. L., Rhoads, J. E., Malhotra, S.,  Pirzkal, N.,  Wang, J. X. 2009, ApJ, 660, 1023
\bibitem[\protect\citeauthoryear{Finkelstein et al.}{2008}]{Fin2008} Finkelstein S.~L., Rhoads J.~E., Malhotra S., Grogin N., Wang J., 2008, ApJ, 678, 655 
\bibitem[\protect\citeauthoryear{Finkelstein et al.}{2009a}]{Fin2009a} Finkelstein S.~L., Cohen S.~H., Malhotra S., Rhoads J.~E., 2009, ApJ, 700, 276 
\bibitem[\protect\citeauthoryear{Finkelstein et al.}{2009b}]{Fin2009b} Finkelstein, S. L., Rhoads, J. E., Malhotra, S., Grogin, N.  2009, ApJ, 691, 465
\bibitem[\protect\citeauthoryear{Finkelstein et al.}{2009c}]{Fin2009c} Finkelstein S.~L., Cohen S.~H., Malhotra S., Rhoads J.~E., Papovich C., Zheng Z.~Y., Wang J.-X., 2009, ApJ, 703, L162 
\bibitem[\protect\citeauthoryear{Fioc \& Rocca-Volmerange.}{1997}]{PEGASE}  Fioc, M., Rocca-Volmerange, B., 1997, A\&A, 326, 950
\bibitem[\protect\citeauthoryear{Frye et al.}{2007}]{Frye2007} Frye, B. L., et al. 2007, ApJ, 665, 921
\bibitem[\protect\citeauthoryear{Gawiser et al.}{2006}]{Ga2006} Gawiser, E. et al. 2006, ApJ, 642, L13
\bibitem[\protect\citeauthoryear{Gawiser et al.}{2007}]{Ga2007} Gawiser, E. et al. 2006, ApJ, 671, 278
\bibitem[\protect\citeauthoryear{Haardt, \& Madau.}{2001}]{Haardt2001} Haardt F.,Madau P., 2001, in Neumann D.M., Tran J. T. V., eds, Clusters of Galaxies and the High Redshift Universe Observed in X-rays. Editions Frontieres, Paris.
\bibitem[\protect\citeauthoryear{Haiman, Spaans, \& Quataert.}{2000}]{Haiman00} Haiman Z., Spaans M., Quataert E., 2000, ApJ, 537, L5 
\bibitem[\protect\citeauthoryear{Hansen, \& Oh.}{2006}]{HansenOh2006} Hansen, M., Oh, S. P., 2006, MNRAS, 367, 979
\bibitem[\protect\citeauthoryear{Hayashino et al.}{2004}]{Ha2004} Hayashino, T. et al., 2004, AJ, 245, 208
\bibitem[\protect\citeauthoryear{Hayes et al.}{2010}]{Hayes10} Hayes, Matthew., et al,  2010, arXiv:1010.4796
\bibitem[\protect\citeauthoryear{Hu et al.}{1998}]{Hu98}  Hu, E. M., Cowie, L. L., McMahon, R. G., 1998, ApJ, 502, L99
\bibitem[\protect\citeauthoryear{Hu et al.}{1999}]{Hu99}  Hu, E. M., McMahon, R. G., Cowie, L. L., 1999, ApJ, 522, L9
\bibitem[\protect\citeauthoryear{Hu et al.}{2002}]{Hu02}  Hu, E. M., Cowie, L. L., McMahon, R. G., Capak, P., Iwamuro, F., Kneib, J.-P., Maihara, T., Motohara K., 2002, ApJ, 568, L75
\bibitem[\protect\citeauthoryear{Iye et al.}{2008}]{Iye06} Iye, M., et al. 2006, Nature, 443, 186
\bibitem[\protect\citeauthoryear{Keel et al.}{2005}]{Keel2005} Keel, W. C. 2005, AJ, 129, 1863
\bibitem[\protect\citeauthoryear{Kobayashi, Totani, \& Nagashima.}{2007}]{Kobayashi2007} Kobayashi, A. R. M., Totani, T., Nagashima, M. 2007, ApJ, 670, 919
\bibitem[\protect\citeauthoryear{Kobayashi, Totani, \& Nagashima.}{2010}]{Kobayashi2010} Kobayashi, A. R. M., Totani, T., Nagashima, M. 2010, ApJ, 708, 1119
\bibitem[\protect\citeauthoryear{Kodaira et al.}{2003}]{Kodaira03} Kodaira K., et al., 2003, PASJ, 55, L17
\bibitem[\protect\citeauthoryear{Kunth et al.}{1998}]{Kunth1998} Kunth, D., Mas-Hesse, J. M., Terlevich, E., Terlevich, R., Lequeux, J., Fall, S. M. 1998, A\&A, 334, 11
\bibitem[\protect\citeauthoryear{Kunth et al.}{2003}]{Kunth2003} Kunth, D., Leitherer, C., Mas-Hesse, J. M., \"Ostlin, G., Petrosian, A. 2003, ApJ, 597, 263
\bibitem[\protect\citeauthoryear{Lai et al.}{2008}]{Lai2008} Lai, K. et al. 2008, ApJ, 674, 70
\bibitem[\protect\citeauthoryear{Lequeux et al.}{1995}]{Lequeux1995} Lequeux, J., Kunth, D., Mas-Hesse, J. M., Sargent, W. L. W. 1995, A\&A, 301, 18
\bibitem[\protect\citeauthoryear{Madau.}{1995}]{Madau1995} Madau, P., 1995, ApJ, 441, 18
\bibitem[\protect\citeauthoryear{Marigo.}{2001}]{Marigo2001} Marigo P., 2001, A\&A, 370, 194
\bibitem[\protect\citeauthoryear{Martin.}{2005}]{Martin2005} Martin C. L., 2005, ApJ, 621, 227
\bibitem[\protect\citeauthoryear{Mas-Hesse et al.}{2003}]{Mas-Hesse2003} Mas-Hesse, J. M., Kunth, D., Tenorio-Tagle, G., Leitherer, C., Terlevich, R. J., Terlevich, E. 2003, ApJ, 598, 858
\bibitem[\protect\citeauthoryear{Matsuda et al.}{2004}]{Matsuda04} Matsuda Y., et al., 2004, AJ, 128, 569 
\bibitem[\protect\citeauthoryear{Matsuda et al.}{2005}]{Matsuda05} Matsuda Y., et al., 2005, ApJ, 634, L125 
\bibitem[\protect\citeauthoryear{Matsuda et al.}{2007}]{Matsuda2007}  Matsuda et al., 2007, ApJ, 667, 667
\bibitem[\protect\citeauthoryear{McLinden et al.}{2010}]{McLinden2010} McLinden, E. M., et al. 2010, AJ submitted (arXiv:1006.1895)
\bibitem[\protect\citeauthoryear{Mori, \& Umemura.}{2006}]{MU2006} Mori, M., Umemura, M., 2006, Nature, 440, 644
\bibitem[\protect\citeauthoryear{Mori, Umemura, \& Ferrara.}{2004}]{MUF04} Mori M., Umemura M., Ferrara A., 2004, ApJ, 613, L97 
\bibitem[\protect\citeauthoryear{Nagamine et al.}{2010}]{Nagamine10} Nagamine, K., Ouchi, M., Springel, V., Hernquist, L. 2010, PASJ, 62, 1455
\bibitem[\protect\citeauthoryear{Neufeld.}{1991}]{Neufeld1991}  Neufeld, D. A., 1991, ApJ, 370, 85
\bibitem[\protect\citeauthoryear{Nilsson et al.}{2009}]{Nilsson2009}  Nilsson et al., 2009, A\&A, 498, 13
\bibitem[\protect\citeauthoryear{Nozawa et al.}{2003}]{Nozawa2003} Nozawa T., Kozasa T., Umeda H., Maeda K., Nomoto K., 2003, ApJ, 598, 785
\bibitem[\protect\citeauthoryear{Okamoto et al.}{2005}]{Okamoto2005} Okamoto T., Eke V. R., Frenk C. S., Jenkins A., 2005, MNRAS, 363, 1299
\bibitem[\protect\citeauthoryear{Okamoto et al.}{2010}]{Okamoto2010} Okamoto, T., Frenk, C. S., Jenkins, A., Theuns, T. 2010, MNRAS, 406, 208
\bibitem[\protect\citeauthoryear{Okamoto, \& Frenk.}{2009}]{Okamoto2009} Okamoto T., Frenk C. S., 2009, MNRAS, 399, L174
\bibitem[\protect\citeauthoryear{Okamoto, Nemmen \& Bower.}{2008}]{Okamoto2008} Okamoto T., Nemmen R. S., Bower R. G., 2008b, MNRAS, 385, 161
\bibitem[\protect\citeauthoryear{Osterbrock.}{1989}]{Osterbrock1989} Osterbrock D.E. 1989, Astrophysics of Gaseous Nebulae and Active Galactic Nuclei, University Science Books
\bibitem[\protect\citeauthoryear{Ono et al.}{2010}]{Ono2010} Ono et al., 2010, MNRAS, 402, 1580
\bibitem[\protect\citeauthoryear{Ouchi et al.}{2004}]{Ou04}  Ouchi, M. et al., 2004, ApJ, 611, 660
\bibitem[\protect\citeauthoryear{Ouchi et al.}{2005}]{Ou05}  Ouchi, M. et al., 2005, ApJ, 620, L1
\bibitem[\protect\citeauthoryear{Ouchi et al.}{2008}]{Ouchi08} Ouchi, M., et al. 2008, ApJS, 176, 301
\bibitem[\protect\citeauthoryear{Pentericci et al.}{2007}]{Pentericci2007} Pentericci, L., Grazian, A., Fontana, A., Salimbeni, S., Santini, P., De Santis, C., Gallozzi, S., Giallongo, E. 2007, A\&A, 471, 433
\bibitem[\protect\citeauthoryear{Pettini et al.}{2002}]{Pettini2002} Pettini, M., Rix, S. A., Steidel, C. C., Hunt, M. P., Shapley, A. E., Adelberger, K. L. 2002, Ap\&SS, 281, 461
\bibitem[\protect\citeauthoryear{Portinari, Chiosi, \& Bressan.}{1998}]{Portinari1998} Portinari L., Chiosi C., Bressan A., 1998, A\&A, 334, 505
\bibitem[\protect\citeauthoryear{Schaerer.}{2003}]{Schaerer03} Schaerer, D. 2003, A\&A, 397, 527
\bibitem[\protect\citeauthoryear{Shapley et al.}{2006}]{Shapley2006} Shapley, A. E., Steidel, C. C., Pettini, M., Adelberger,K. L. 2003, ApJ, 588, 65
\bibitem[\protect\citeauthoryear{Shimasaku et al.}{2003}]{Shimasaku03} Shimasaku K., et al., 2003, ApJ, 586, L111 
\bibitem[\protect\citeauthoryear{Shimasaku et al.}{2006}]{Shimasaku06} Shimasaku K., et al., 2006, PASJ, 58, 313 
\bibitem[\protect\citeauthoryear{Shimizu et al.}{2007}]{S2007} Shimizu, I., Umemura, M., Yonehara, A. 2007, MNRAS, 380, 49L
\bibitem[\protect\citeauthoryear{Shimizu et al.}{2010}]{Shimizu2010} Shimizu, I., Umemura, M., 2010, MNRAS, 406, 913
\bibitem[\protect\citeauthoryear{Smail et al.}{2004}]{Smail04} Smail, Ian., Chapman, S. C., Blain, A. W., Ivison, R. J., ApJ, 616, 71
\bibitem[\protect\citeauthoryear{Spergel et al.}{2003}]{WMAP} Spergel, D. N., et al., 2003, ApJS, 148, 175
\bibitem[\protect\citeauthoryear{Springel et al.}{2001}]{Springel2001} Springel V., White S. D. M., Tormen G., Kauffmann G., 2001, MNRAS,
328, 726
\bibitem[\protect\citeauthoryear{Springel.}{2005}]{Gadget} Springel V., 2005, MNRAS, 364, 1105
\bibitem[\protect\citeauthoryear{Swinbank et al.}{2007}]{Swinbank2007} Swinbank, A. M., Bower, R.G., Smith,G. P.,Wilman, R. J., Smail, I., Ellis, R. S., Morris, S. L., Kneib, J.-P. 2007, MNRAS, 376, 479
\bibitem[\protect\citeauthoryear{Tamura et al.}{2009}]{SMGLAE}  Tamura et al., 2009, Nature, 459, 61
\bibitem[\protect\citeauthoryear{Taniguchi et al.}{2005}]{Taniguchi05} Taniguchi Y., et al., 2005, PASJ, 57, 165 
\bibitem[\protect\citeauthoryear{Taniguchi, \& Shioya.}{2000}]{TS00} Taniguchi Y., Shioya Y., 2000, ApJ, 532, L13 
\bibitem[\protect\citeauthoryear{Tapken et al.}{2007}]{Tapken2007} Tapken, C., Appenzeller, I., Noll, S., Richling, S., Heidt, J., Meink\"ohn, E., Mehlert, D. 2007, A\&A, 467, 63
\bibitem[\protect\citeauthoryear{Todini \& Ferarra.}{2001}]{Todini2001} Todini P., Ferarra A., 2001, MNRAS, 325, 726
\bibitem[\protect\citeauthoryear{Uchimoto et al.}{2008}]{Uchimoto2008}  Uchimoto et al., 2008, PASJ, 60, 683
\bibitem[\protect\citeauthoryear{Wiersma, Schaye, \& Smith.}{2009}]{Wiersma2009} Wiersma R. P. C., Schaye J., Smith B. D., 2009, MNRAS, 393, 99
\bibitem[\protect\citeauthoryear{Wilman et al.}{2005}]{Wilman2005} Wilman, R. J., Gerssen, J., Bower, R. G.,Morris, S. L., Bacon, R., de Zeeuw, P. T., Davies, R. L. 2005, Nature, 436, 227
\bibitem[\protect\citeauthoryear{Zheng et al.}{2010a}]{Zheng2010a} Zheng, Z., Cen, R., Trac, H., Miralda-Escud\'e, J., 2010, ApJ, 716, 574
\bibitem[\protect\citeauthoryear{Zheng et al.}{2010b}]{Zheng2010b} Zheng, Z., Cen, R., Weinberg, D., Trac, H., Miralda-Escud\'e, J., 2010 (arXiv:1010.3017)
\end{thebibliography}
\end{document}